\documentclass[epj]{svjour}
\usepackage{graphics}
\usepackage{color}
\usepackage{graphicx}
\usepackage{epstopdf}
\usepackage{subfigure}
\usepackage{amssymb}
\usepackage{CJK}
\usepackage{indentfirst}
\usepackage{amsmath}
\usepackage[square, comma, sort&compress, numbers]{natbib}

\begin{document}

\title{Statistical analysis with cosmic-expansion-rate measurements and two-point diagnostics}

\author{Xiaogang Zheng\inst{1,2,3}, Marek Biesiada\inst{2,3}, Xuheng Ding\inst{1,2}, Shuo Cao\inst{2}\thanks{\emph{e-mail:} caoshuo@bnu.edu.cn}, Sixuan Zhang \inst{2} \and Zong-Hong Zhu\inst{1,2}}

\institute{School of Physics and Technology, Wuhan University, Wuhan 430072, China;
\and Department of Astronomy, Beijing Normal University, Beijing, 100875, China;
\and Department of Astrophysics and Cosmology, Institute of Phyisics, University of Silesia, Uniwersyecka 4, 40-007, Katowice, Poland}

\date{Received: date / Revised version: date}

\abstract{
Direct measurements of Hubble parameters $H(z)$ are very useful for
cosmological model parameters inference. Based on them,
Sahni, Shafieloo and Starobinski introduced a two-point diagnostic
$Omh^2(z_i, z_j)$ as an interesting tool for testing the validity of
the $\Lambda$CDM model. Applying this test they found a tension
between observations and predictions of the $\Lambda$CDM model.
We use the most comprehensive compilation $H(z)$ data from baryon
acoustic oscillations (BAO) and differential ages (DA) of passively
evolving galaxies to study cosmological models using the Hubble
parameters itself and to distinguish whether $\Lambda$CDM model is
consistent with the observational data with statistical analysis of
the corresponding $Omh^2(z_i, z_j)$ two-point diagnostics.
Our results show that presently available $H(z)$ data
significantly improve the constraints on cosmological parameters. 
The corresponding statistical $Omh^2(z_i, z_j)$ two-point 
diagnostics seems to prefer the quintessence
with $w>-1$ over the $\Lambda$CDM model. Better and more accurate
prior knowledge of the Hubble constant, will considerably improve
the performance of the statistical $Omh^2(z_i, z_j)$ method. }
\PACS{{cosmology}{Hubble parameter} \and {methods}{statistical}}

\authorrunning{Xiaogang Zheng, et al.}

\titlerunning{Statistical analysis with $H(z)$ and $Omh^2(z_i,z_j)$ }
\maketitle

\section{Introduction}\label{sec:introduction}

The discovery of accelerating expansion of the Universe
\citep{Riess1998,Perlmutter1999} created a big challenge for the
modern science and stimulated cosmologists to investigate the
essentials of this phenomenon. In order to explain present
acceleration of the Universe, there should exist some mechanism
providing a repulsive effect. There are two broad ways of achieving
this: considering the modified gravity \citep{Clifton2012} or adding
an exotic dark energy component \citep{Frieman2008} to the matter
content of the Universe. The simplest solution along the second line
of reasoning is the $\Lambda$CDM model in which the cosmological
constant $\Lambda$ acts as a repulsive component in addition to
ordinary cold dark matter and - now dynamically unimportant - CMB
radiation or cosmic neutrinos. However, the cosmological constant,
while being the most parsimonious choice is far from being a
satisfactory explanation both theoretically (fine tuning and
coincidence problems) and from the observational point of view
\citep{Buchert16}. Because there is no clear theoretical preference
for the alternative to the $\Lambda$CDM model, it is reasonable to
take a phenomenological approach to parameterize the unknown by
hypothetical fluid with an equation of state $p=w\rho$ where $w$
coefficient might be constant or allowed to vary with cosmic time
$w(z)=w_0+w_a\frac{z}{1+z}$ \citep{Chevalier2001,Linder2003}. Such
models are known as wCDM and CPL, respectively. Standard
$\Lambda$CDM is nested within such classes of models.

The most straightforward technique to constrain cosmological
equation of state is by constructing the Hubble diagram $d_{L,A}(z)$
using either luminosity or angular diameter distances to the objects
whose redshifts are known \citep{Cao12,Cao14,Cao15,Cao17a,Cao17b}.
This approach demands either standard candles like SN Ia or standard
rulers like CMB acoustic peaks or BAO. One should be cautious,
however, about the way they are calibrated in order not to fall into
circularity problems with respect to the cosmological model assumed
during the calibration. From this perspective, another very
attractive probe - Hubble function at different redshifts $H(z)$ -
is becoming accessible. In particular, $H(z)$ measurements from the
so called cosmic chronometers, i.e. differential ages (DA) of
passively evolving galaxies are free from any prior assumption
concerning cosmology, only uncertainty being of astrophysical origin
(the adopted population synthesis model).

Recently, using DA technique, \citet{Moresco2015, Moresco2016a}
provided another few $H(z)$ measurements in addition to already
existing data (see \citet{Ding2015} for the compilation). They also
used the whole compilation of $H(z)$ from DA to constrain cosmology
\citep{Moresco2016b}. Expansion rates at different redshifts not
only allowed to use this pure information for cosmographic purposes
but opened also a new chapter in using the so called $Om(z)$
diagnostics. They were introduced by \cite{Sahni2008} in order to
distinguish between $\Lambda$CDM and other dark energy scenarios.
This diagnostics is defined as
\begin{equation} \label{eqOmz}
Om(z)\equiv\frac{{E}^2(z)-1}{(1+z)^3-1}
\end{equation}
where ${E}(z) \equiv H(z)/H_0$ is the dimensionless expansion rate
and in the $\Lambda$CDM model it should be equal exactly to the
present value of matter density $Om(z)=\Omega_{m,0}$. Its advantage
as a screening test for the $\Lambda$CDM (formal function of the
redshift $Om(z)$ should be just a constant) is clear. Later on, they
developed it further by introducing a two-point diagnostic
$Omh^2(z_i,z_j)$ \citep{Sahni2012}
\begin{equation} \label{eqOmh2}
Omh^2(z_i,z_j)=\frac{h^2(z_i)-h^2(z_j)}{(1+z_i)^3-(1+z_j)^3}
\end{equation}
where ${h}(z) \equiv H(z)/100$, and subsequently used it in
\cite{Sahni2014} to perform this test on three accurately measured
values of $H(z)$ from BAO demonstrating a tension with the value of
$\Omega_{m,0}h^2$ given by \citet{Planck2014}. Later,
\citet{Ding2015} collected a larger $H(z)$ sample (6 from BAO
measurements and 23 from DA measurements) to do this test confirming
that the tension exists. The two-point diagnostics has an advantage
that if we know Hubble parameters at $n$ different redshifts, we can
get $n(n-1)/2$ pairs of data. This enlargement of statistical sample
for inference occurs at the expense of non-trivial statistical
properties of observables \citep{Zheng16}.

As already mentioned, $H(z)$ can be used as a cosmological probe to
constrain cosmological parameters directly
\citep{Cao11a,Cao11b,Cao13,Chen15}. However, it is also tempting to
perform the fit cosmological parameters based on the two-point
diagnostics. Therefore, in this paper we constrain the cosmological
models not only using $H(z)$ directly, but also using the two-point
$Omh^2(z_i,z_j)$ diagnostic. The rest of the paper is organized as
follows. In Section~\ref{sec:daandme}, we briefly introduce the
observational Hubble parameters, and present our methodology to
constrain cosmology with $Omh^2(z_i,z_j)$ probe. We show our results
followed by discussion in Section~\ref{sec:reanddi}. Finally, we
conclude in Section~\ref{sec:conclusion}.

\section{Data and Method}\label{sec:daandme}
\subsection{Empirical $H(z)$ data and constraints based directly on them }

\begin{figure}[htbp]
\begin{center}
\centering
\includegraphics[angle=0,width=75mm]{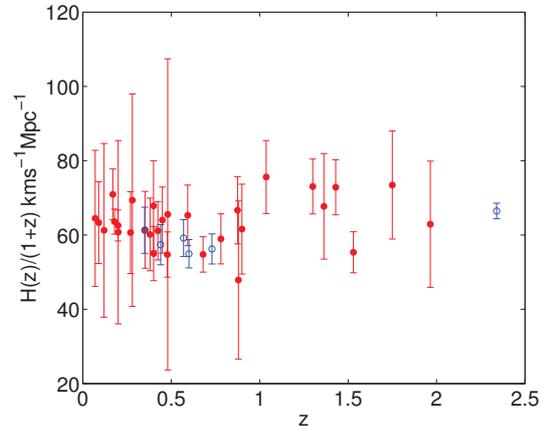}
\caption{\label{fig1} Observed $H(z)$ data from DA (red dot) and BAO
(blue circle) with corresponding uncertainties.}
\end{center}
\end{figure}

We used a collection of totally 36 measurements of $H(z)$ shown in
Fig.~(\ref{fig1}). Among them, 30 data points come from cosmic
chronometers \citep{Jimenez2003, Simon2005, Stern2010, Moresco2012,
Zhang2014, Moresco2015, Moresco2016a}, i.e. the differential ages of
passively evolving galaxies as a function of redshift. Other 6
points come from the BAO peak position as a standard ruler in the
radial direction \citep{Chuang2013, Blake2012, Anderson2013,
Delubac2015}. Because the $H(z)$ data come from two different
techniques, and moreover one of the BAO points -- the one at the
highest redshift \citep{Delubac2015} -- was obtained in a different
way than other BAO data (from the $Ly\alpha$ forest) we also divided
our data set (full $n=36$ sample) into sub-samples: $n=35$ points --
high $z$ BAO excluded, $n=30$ -- from cosmic chronometers (DA) only
and $n=6$ from BAO only. Such division is dictated by desire to
reveal possible systematics due to inhomogeneous sample.

\begin{figure*}
\begin{center}
\centering
\includegraphics[angle=0,width=52mm]{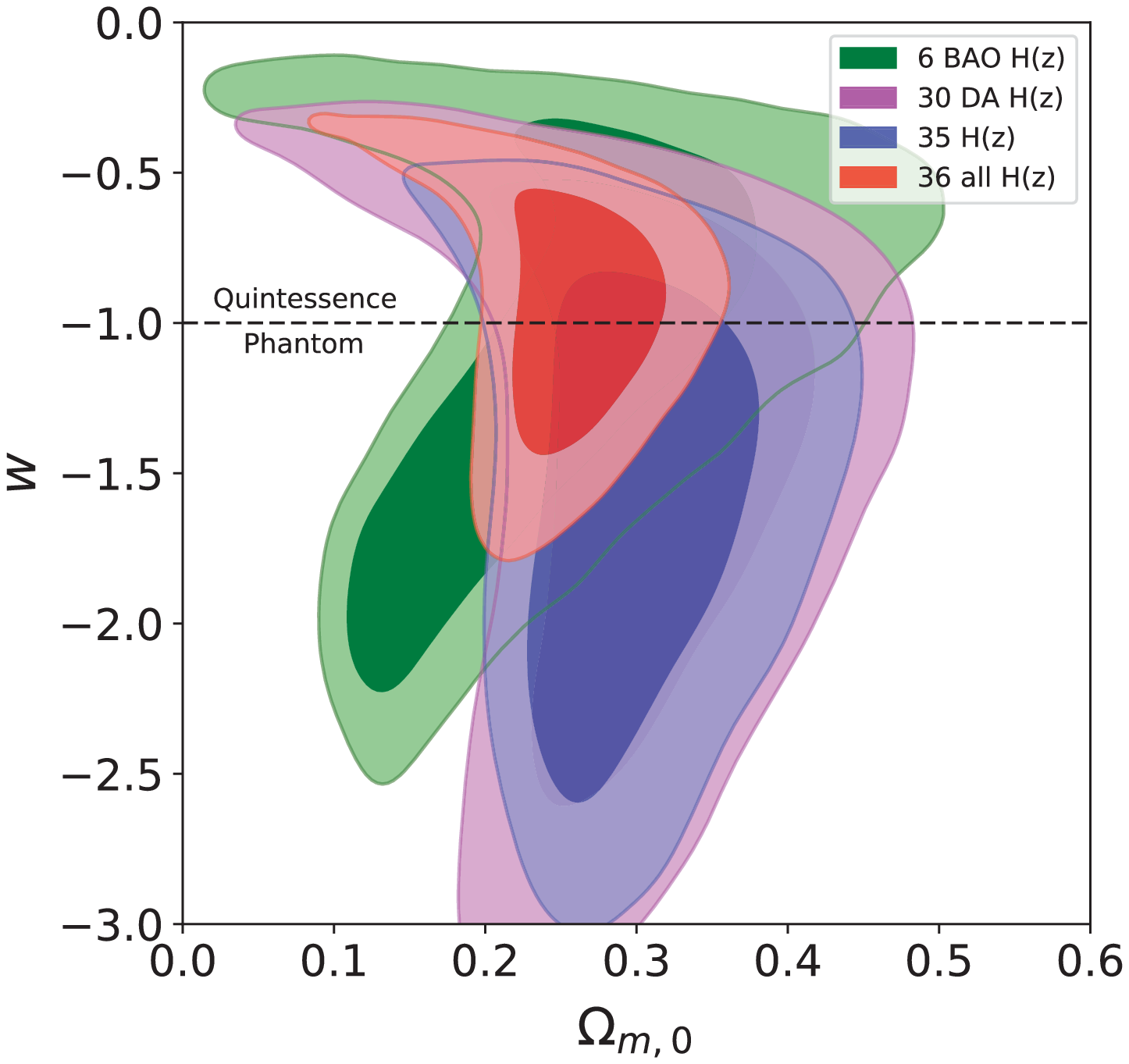}
\includegraphics[angle=0,width=50mm]{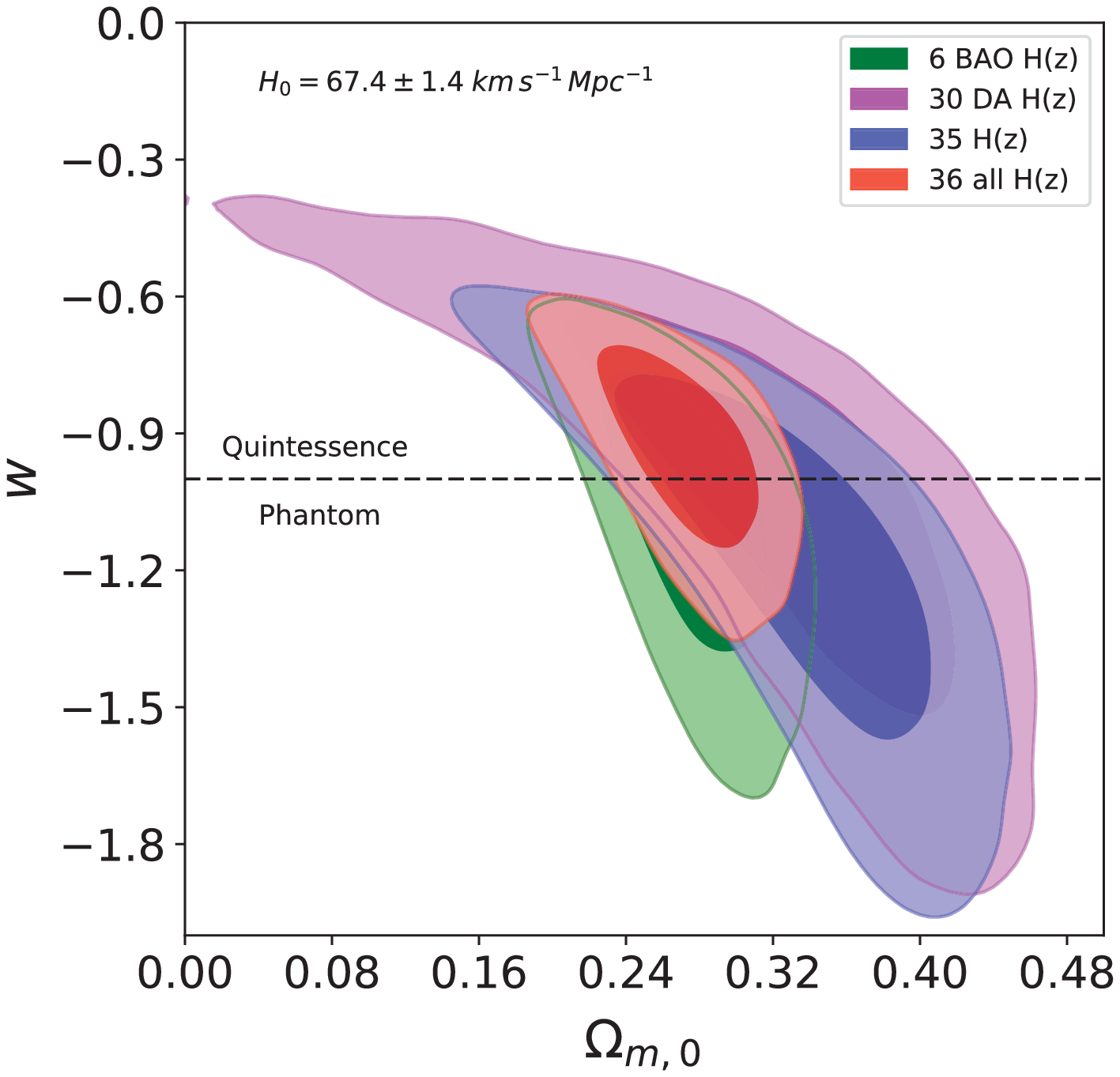} \includegraphics[angle=0,width=50mm]{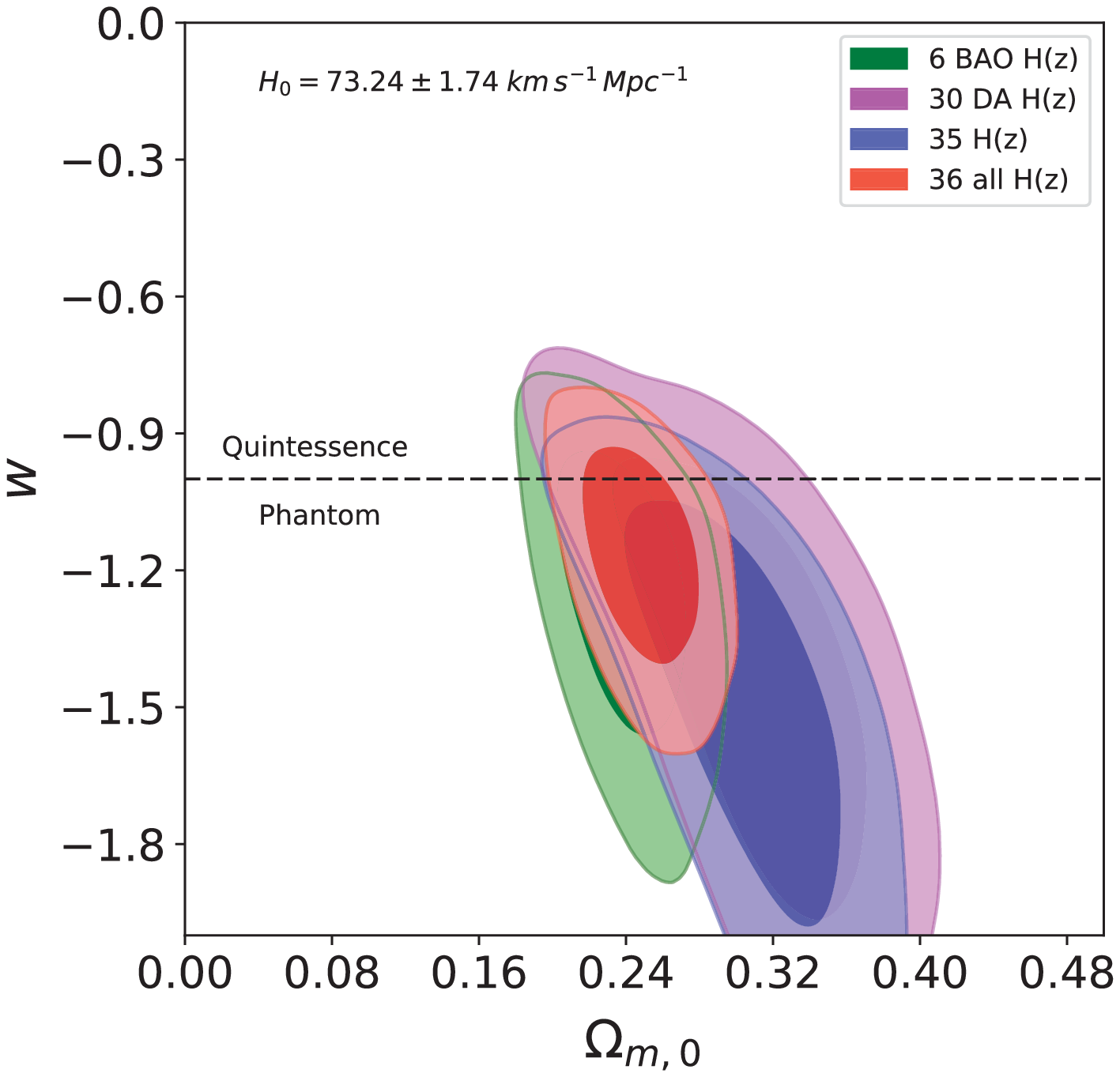}
\caption{\label{fig6} Constraints on the parameters of wCDM
cosmological model obtained with the expansion rate measurements
$H(z)$. Upper plot was obtained using the reduced chi-square
function Eq.(\ref{eqchi_Hmin}). Lower left one with Gaussian prior
$H_0=67.4\pm1.4\;km\,s^{-1}\,Mpc^{-1}$ from \textit{Planck} result
\citep{Planck2014} and lower right one with prior
$H_0=73.24\pm1.74\;km\,s^{-1}\,Mpc^{-1}$ from local measurements
\citep{Riess2016}. }
\end{center}
\end{figure*}

\begin{figure*}
\begin{center}
\centering
\includegraphics[angle=0,width=65mm]{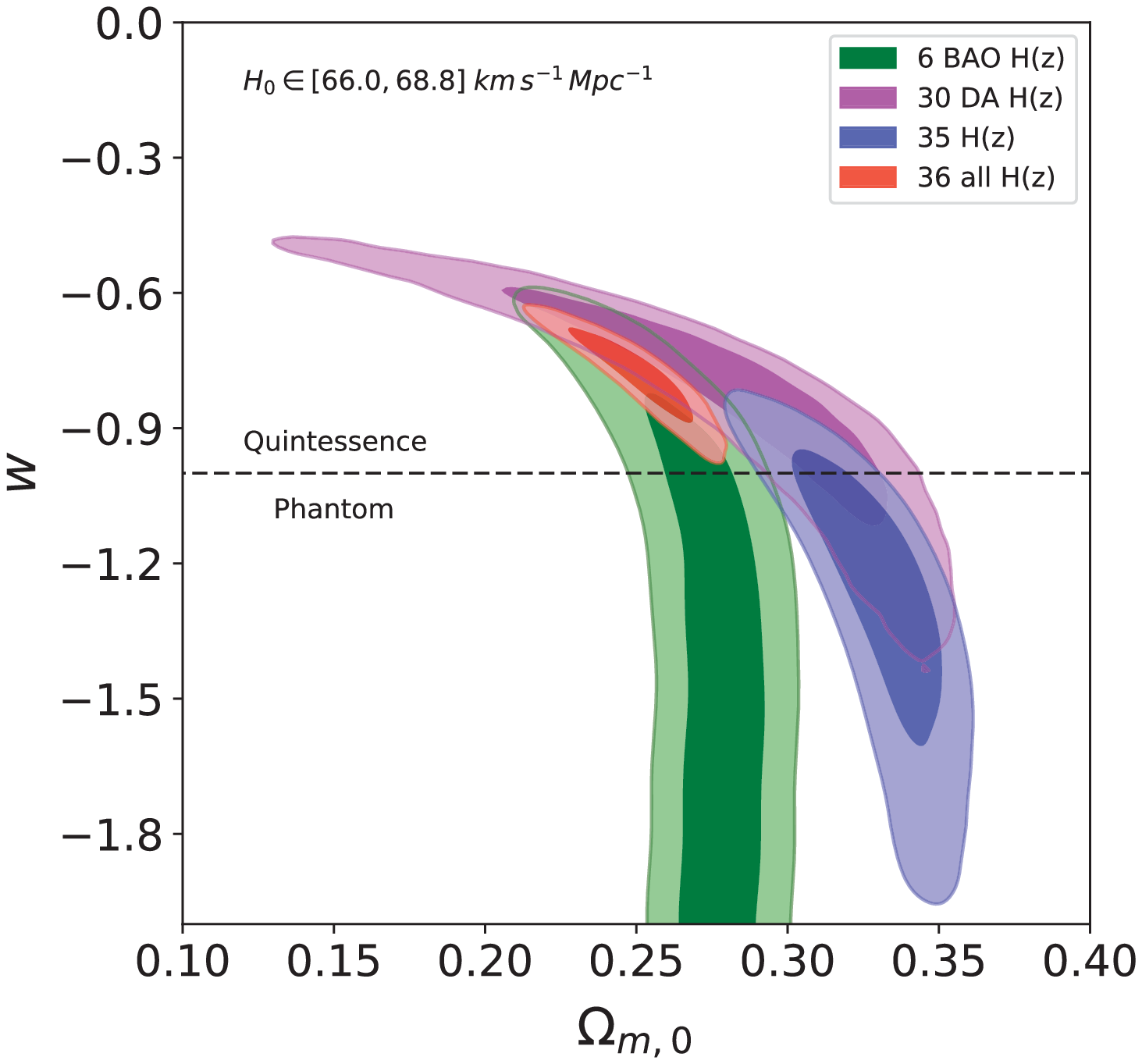}
\includegraphics[angle=0,width=65mm]{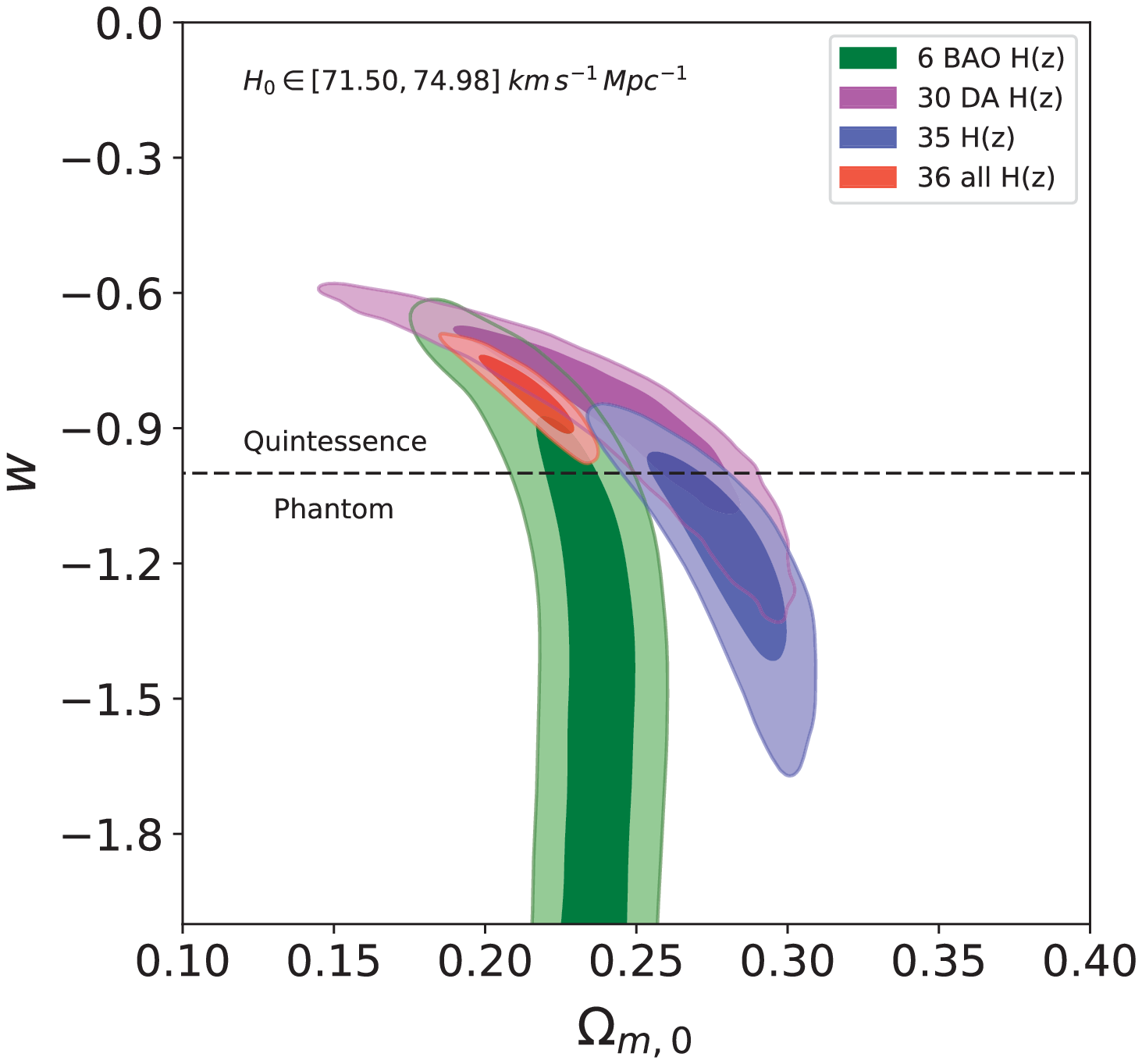}
\caption{\label{fig7} Constraints on the parameters of wCDM
cosmological model obtained using $Omh^2(z_i,z_j)$ two-point
diagnostics. Left panel: with Uniform prior $H_0\in[66.0,
68.8]\;km\,s^{-1}\,Mpc^{-1}$ corresponding to \textit{Planck} result
\citep{Planck2014}; Right panel: with Uniform prior $H_0\in[71.50,
74.98]\;km\,s^{-1}\,Mpc^{-1}$ corresponding to the results from
\citet{Riess2016}.}
\end{center}
\end{figure*}

\begin{figure*}
\begin{center}
\centering
\includegraphics[angle=0,width=60mm]{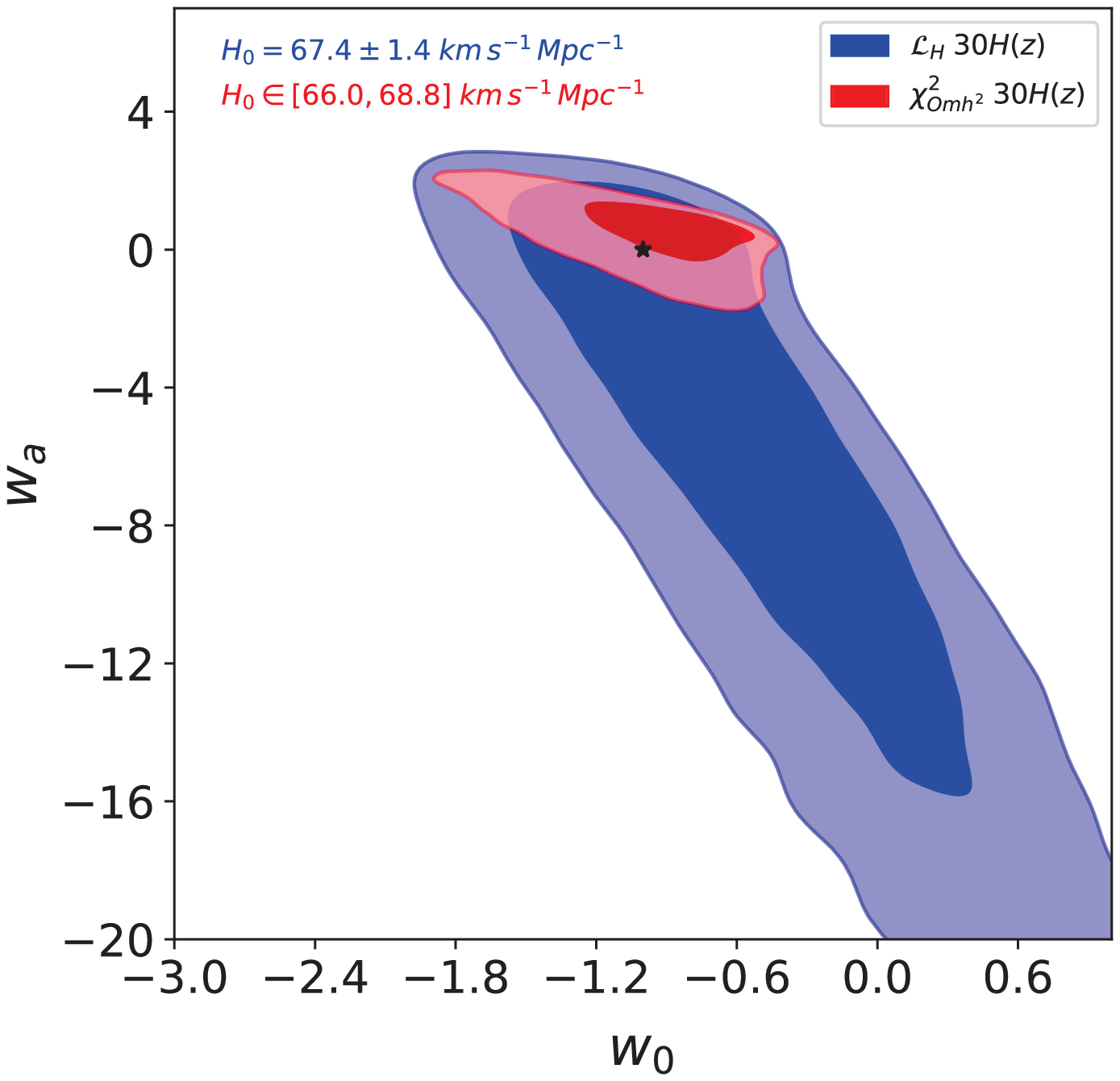} \includegraphics[angle=0,width=62mm]{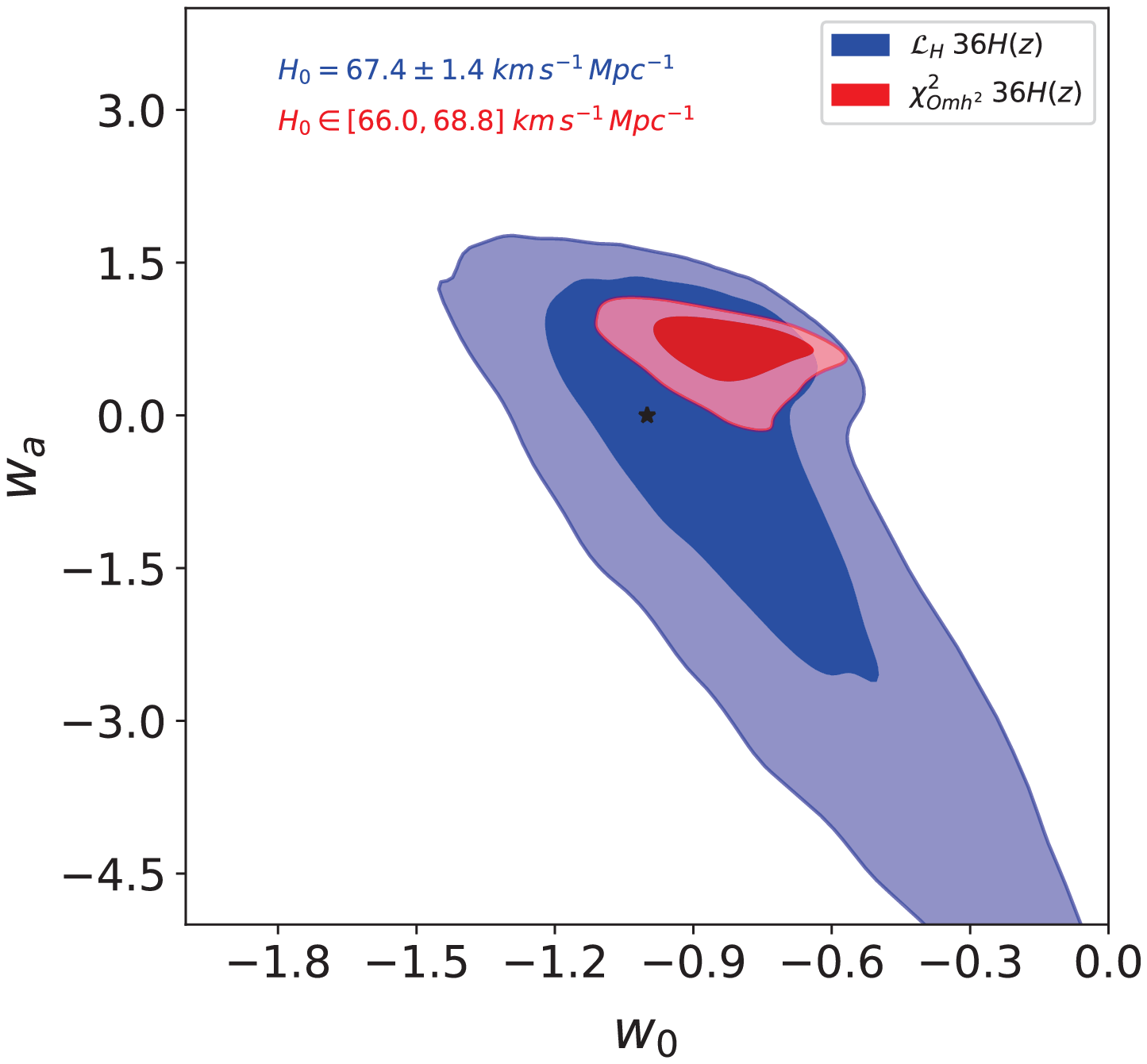}
\includegraphics[angle=0,width=60mm]{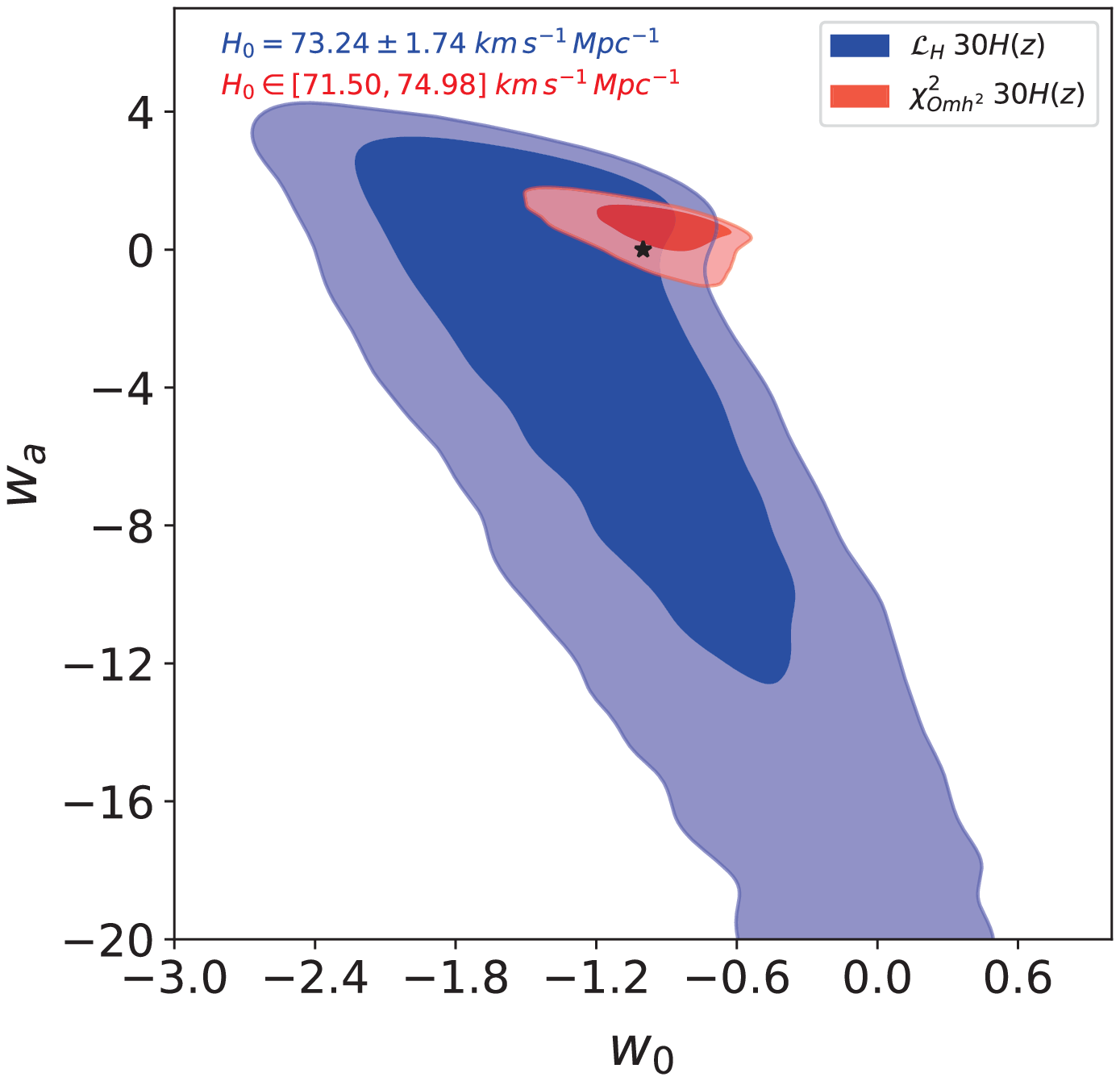} \includegraphics[angle=0,width=62mm]{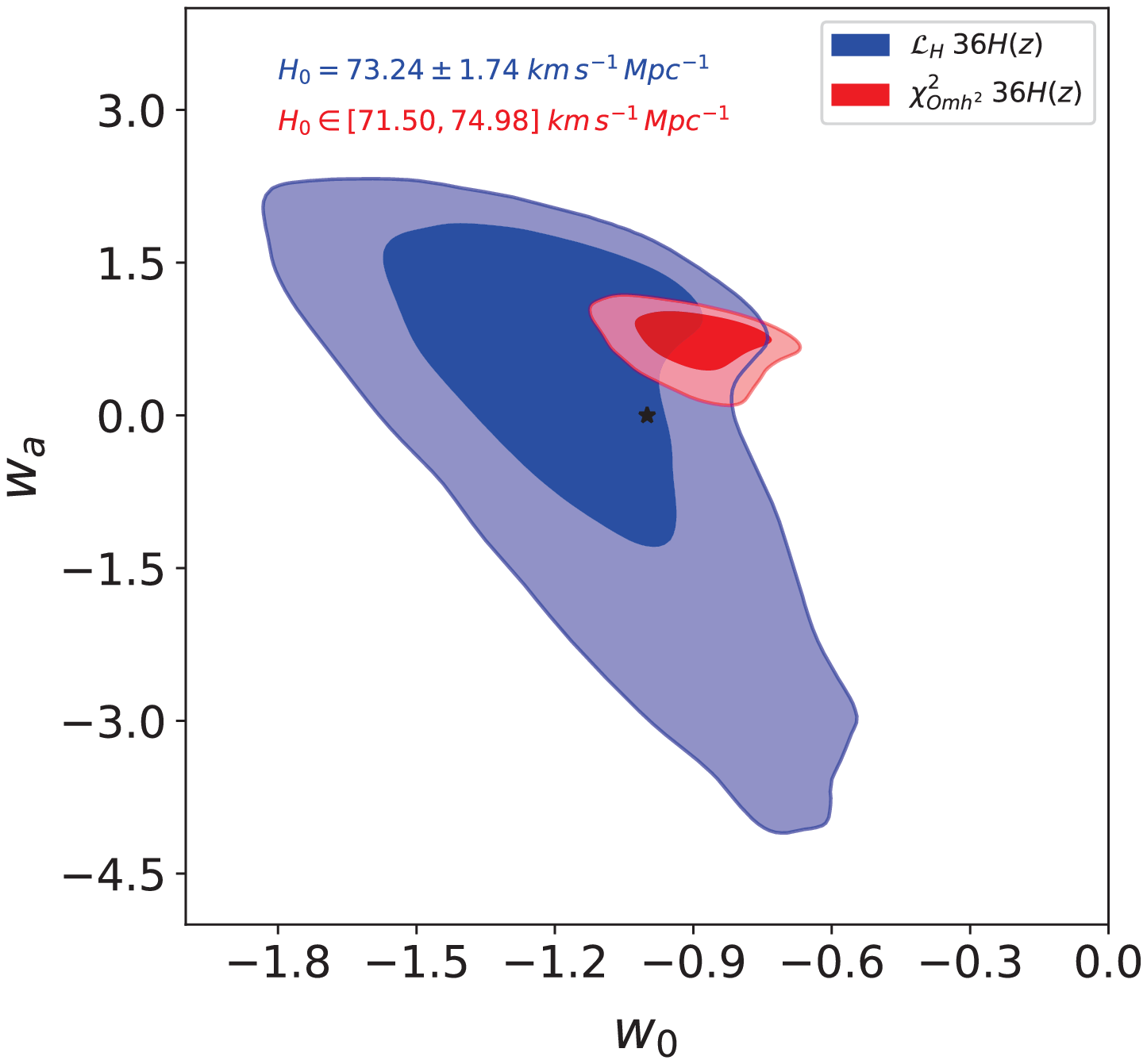}
\caption{\label{fig8} Constraints on the CPL parameters from
 30 DA H(z) data points and 36 total sample. Blue contour: Hubble parameters constraint with $H_0$ Gaussian priors; Red contour: $Omh^2(z_i,z_j)$ method constraint with $H_0$ Uniform priors.
Left two panels assume $H_0$ prior from Planck
result\citep{Planck2014} and the right two panels correspond to the
$H_0$ from \citet{Riess2016}. }
\end{center}
\end{figure*}

We will use these data to estimate cosmological parameters denoted
in short as $\textbf{p}$. In particular,
$\textbf{p}=\{\Omega_{m,0},w\}$ for wCDM and
$\textbf{p}=\{\Omega_{m,0},w_0,w_a\}$ for CPL model. It is obvious
that $\Lambda$CDM model with $\textbf{p}=\{\Omega_{m,0}\}$ is nested
within the above mentioned models and is equivalent to wCDM with $w$
parameter fixed at $w=-1$ or CPL with $w_0 = -1$ and $w_a = 0$
fixed, so in this case $\textbf{p}={\Omega_{m,0}}$. For completeness
and cross-checks we will also report fits on the present matter
density parameter in $\Lambda$CDM model. Let us note that we do not
consider the Hubble constant $H_0$ as a free parameter for fitting.
Therefore, as described in details below, we either marginalize over
$H_0$ (in some specific way) or use an informative prior for it. In
order to estimate the best fitted values of these parameters we will
maximize the likelihood derived from the $\chi^2$ function. In the
case of constraints based on $H(z)$ data it reads:
\begin{equation}\label{eqchi_H}
\chi^2_{H(z)}(H_0,\textbf{p})=\sum^{n}_{i=1}\left[\frac{H(z_i;H_0,\textbf{p})_{th}-H(z_i)_{obs}}{\sigma_{H(z_i)_{obs}}}\right]^2
\end{equation}
Because we treat $H_0$ as a nuisance parameter, one can factor it
out: $H(z;H_0,\textbf{p})_{th}=H_0E(z;\textbf{p})$ and rewrite
Eq.~(\ref{eqchi_H}) in the following way
\begin{align}\label{eqchi_Hp}
& \chi^2_{H(z)}(H_0,\textbf{p})=\sum^{n}_{i=1}\left[\frac{H_0E(z_i;\textbf{p})-H(z_i)_{obs}}{\sigma_{H(z_i)_{obs}}}\right]^2 \nonumber \\
&
=H_0^2\sum^{n}_{i=1}\frac{E^2(z_i;\textbf{p})}{\sigma^2_{H(z_i)_{obs}}}
 +\sum^{n}_{i=1}\frac{H(z_i)^2_{obs}}{\sigma^2_{H(z_i)_{obs}}} \nonumber \\
&
-2H_0\sum^{n}_{i=1}\frac{H(z_i)_{obs}E(z_i;\textbf{p})}{\sigma^2_{H(z_i)_{obs}}}
\end{align}
where only the dimensionless expansion rate depends explicitly on
the cosmological model parameters. Let us recall that in the wCDM
model with constant $w$ coefficient in the equation of state it
reads:
\begin{align}\label{eqEzwCDM}
& E(z;w,\Omega_{m,0})= \nonumber \\
&
\left(\Omega_{m,0}(1+z)^3+(1-\Omega_{m,0})(1+z)^{3(1+w)}\right)^{1/2}
\end{align}
while and for the Chevalier-Polarski-Linder (CPL) parametrization
\citep{Chevalier2001,Linder2003}, one has:
\begin{align}\label{eqEzCPL}
& E(z;w,\Omega_{m,0})= \nonumber \\
&
\left(\Omega_{m,0}(1+z)^3+(1-\Omega_{m,0})(1+z)^{3(1+w_0+w_a)}e^{(-\frac{3w_az}{1+z})}\right)^{1/2}
\end{align}

Introducing auxiliary quantities:
\begin{align}\label{eqH_ABC}
& Q_1=\sum^n_{i=1}\frac{E^2(z_i;\textbf{p})}{\sigma^2_{H(z_i)_{obs}}} \nonumber \\
& Q_2=\sum^n_{i=1}\frac{H_{obs}(z_i)E(z_i;\textbf{p})}{\sigma^2_{H(z_i)_{obs}}} \nonumber \\
& Q_3=\sum^n_{i=1}\frac{H^2_{obs}(z_i)}{\sigma^2_{H(z_i)_{obs}}}
\end{align}
one can rewrite Eq.~(\ref{eqchi_Hp}) as
\begin{equation}\label{eqchi_Hs}
\chi^2_{H(z)}(H_0,\textbf{p})=Q_1H_0^2-2Q_2H_0+Q_3
\end{equation}
Now, it is easy to see that the reduced chi-square minimized with
respect to the nuisance parameter $H_0$ is equal to
\begin{equation}\label{eqchi_Hmin}
\chi^2_{H(z)}(\textbf{p})=Q_3-\frac{Q_2^2}{Q_1}
\end{equation}
at $H_0=Q_2/Q_1$ and one can use it further to constrain parameters
$\textbf{p}$ without any prior assumptions about $H_0$. This
approach is alternative to standard procedure of marginalizing over
$H_0$.

Another approach is to take an informative prior for $H_0$.
Following \citet{Farooq2013}, we will assume that the prior
distribution of $H_0$ is Gaussian with the mean $\bar{H_0}$ and the
standard deviation $\sigma_{H_0}$:
\begin{equation}\label{eqH0PDF}
P(H_0)=\frac{1}{\sqrt{2\pi\sigma^2_{H_0}}}e^{-(H_0-\bar{H_0})^2/(2\sigma^2_{H_0})}
\end{equation}
Then, we can build the posterior likelihood function
$\mathcal{L}_H(\textbf{p})$ by marginalizing over $H_0$
\begin{equation}\label{eqHzL}
\mathcal{L}_{H}(\textbf{p})=\int^{\infty}_0e^{-\chi^2_{H(z)}(H_0,\textbf{p})}P(H_0)dH_0
\end{equation}
Introducing
\begin{align}\label{eqHdefine}
& \alpha=\frac{1}{\sigma^2_{H_0}}+Q_1; \nonumber \\
& \beta=\frac{\bar{H_0}}{\sigma^2_{H_0}}+Q_2; \nonumber \\
& \gamma=\frac{\bar{H_0}^2}{\sigma^2_{H_0}}+Q_3
\end{align}
where the terms $Q_1,Q_2,Q_3$ are the same as in
Eq.~(\ref{eqH_ABC}), and performing the integral analytically one
arrives at the following expression for the posterior likelihood:
\begin{equation}\label{eqHzLL}
\mathcal{L}_{H}(\textbf{p})=\frac{1}{2\sqrt{\alpha\sigma^2_{H_0}}}e^{\left[-\frac{1}{2}(\gamma-\frac{\beta^2}{\alpha})\right]}\left[1+erf(\frac{\beta}{\sqrt{2\alpha}})\right]
\end{equation}
where $erf(x)=\frac{2}{\sqrt{\pi}}\int^x_0e^{-t^2}dt$. Details of
the derivation can be found in \citet{Farooq2013}.

Then, we maximize the likelihood $\mathcal{L}_{H}(\textbf{p})$, with
respect to the parameters $\textbf{p}$ in order to find the
best-fitted parameter values $\textbf{p}_0$.

\begin{table}[htp]
\caption{ Best fitted parameters in $\Lambda$CDM cosmological model
using $H(z)$ data alone and $Omh^2(z_i,z_j)$ two-point diagnostics.
Fits done on different sub-samples are reported. First panel
corresponds to the reduced $\chi^2$ method. Second and third panel
correspond to priors on $H_0$ taken after \textit{Planck}
\citep{Planck2014}
and after \citet{Riess2016}. 
}
\begin{center}
{{\scriptsize
 \begin{tabular}{l c c} \hline\hline
   $\chi^2_{H(z)}(\textbf{p})$  & $\Omega_{m,0}$ &  $\chi^2_{d.o.f}$ \\ \cline{2-3}
   reduced chi-square & & \\
$n=6$ BAO H(z)  & $0.28^{+0.09}_{-0.07}$ &  $0.64/5$ \\
$n=30$ DA H(z)  & $0.32^{+0.10}_{-0.08}$ &  $14.50/29$ \\
$n=35$ H(z)     & $0.31^{+0.09}_{-0.08}$ &  $16.84/34$ \\
$n=36$ all H(z) & $0.26^{+0.05}_{-0.04}$ &  $17.78/35$ \\
\hline \hline
  $\mathcal{L}_H(\textbf{p})$  & $\Omega_{m,0}$ &  $\chi^2_{d.o.f}$ \\ \cline{2-3}
Gaussian Prior & $H_0=67.4\pm1.4$ km/s/Mpc  & \\
$n=6$ BAO H(z)  & $0.270^{+0.034}_{-0.032}$ &  $1.22/5$ \\
$n=30$ DA H(z)  & $0.329^{+0.062}_{-0.053}$ &  $15.14/29$ \\
$n=35$ H(z)     & $0.310^{+0.051}_{-0.048}$ & $16.62/34$ \\
$n=36$ all H(z) & $0.281^{+0.032}_{-0.031}$ &  $19.17/35$ \\
\cline{2-3}
Gaussian Prior & $H_0=73.24\pm1.74 $ km/s/Mpc  & \\
$n=6$ BAO H(z)  & $0.224^{+0.032}_{-0.029}$ &  $2.80/5$   \\
$n=30$ DA H(z)  & $0.261^{+0.056}_{-0.047}$ &  $17.38/29$ \\
$n=35$ H(z)     & $0.241^{+0.050}_{-0.042}$ &  $20.45/34$ \\
$n=36$ all H(z) & $0.238^{+0.031}_{-0.028}$ &  $20.70/35$ \\
 \hline \hline
   $\chi^2_{Omh^2}(H_0,\textbf{p})$  & $\Omega_{m,0}$ &  $\chi^2_{d.o.f}$ \\ \cline{2-3}
Uniform Prior &  $H_0\in[66.0, 68.8]$ km/s/Mpc  & \\
$n=6$ BAO H(z)  & $0.272^{+0.018}_{-0.017}$ &  $1.62/13$    \\
$n=30$ DA H(z)  & $0.318^{+0.019}_{-0.019}$ &  $220.46/433$ \\
$n=35$ H(z)     & $0.311^{+0.017}_{-0.016}$ &  $296.21/593$ \\
$n=36$ all H(z) & $0.279^{+0.011}_{-0.011}$ &  $323.33/628$ \\
 \cline{2-3}
Uniform Prior &  $H_0\in[71.50, 74.98]$ km/s/Mpc  & \\
$n=6$ BAO H(z)  & $0.229^{+0.016}_{-0.015}$ &  $1.62/13$ \\
$n=30$ DA H(z)  & $0.268^{+0.017}_{-0.017}$ &  $220.46/433$ \\
$n=35$ H(z)     & $0.261^{+0.016}_{-0.015}$ &  $296.21/593$ \\
$n=36$ all H(z) & $0.234^{+0.012}_{-0.011}$ &  $323.33/628$ \\
 \hline \hline
\end{tabular}} \label{table1}}
\end{center}
\end{table}

\subsection{Constraints based on two-point diagnostics}

So far, the two point diagnostic has been mostly used to test the
validity of $\Lambda$CDM model and to some extent its
generalizations \citep{Sahni2014,Ding2015,Zheng16}. Here, we will
use the $Omh^2(z_i,z_j)$ function for the purpose of constraining
cosmological parameters $\textbf{p}$ following the similar strategy
as described above for expansion rates alone.

Introducing the simplifying notation: $h(z)=H(z)/100$ and
$e(z)=E(z)/100$, one can express theoretically expected
$Omh^2(z_i,z_j;H_0,\textbf{p})_{th}$ and observed
$Omh^2(z_i,z_j)_{obs}$ two point diagnostics as
\begin{align}\label{eqOmh2_thobs}
& Omh^2(z_i,z_j;H_0,\textbf{p})_{th}=\frac{[H_0e(z_i)]^2-[H_0e(z_j)]^2}{(1+z_i)^3-(1+z_j)^3} \nonumber \\
&
Omh^2(z_i,z_j)_{obs}=\frac{[h(z_i)]^2-[h(z_j)]^2}{(1+z_i)^3-(1+z_j)^3}
\end{align}
The $\chi^2$ function for the $Omh^2(z_i,z_j)$ two point diagnostics
is
\begin{align}\label{eqchi_Omh2}
& \chi^2_{Omh^2}(H_0,\textbf{p})= \nonumber \\
&
\sum^{n-1}_{i=1}\sum^{n}_{j=i+1}\left[\frac{Omh^2(z_i,z_j;H_0,\textbf{p})_{th}-Omh^2(z_i,z_j)_{obs}}
{\sigma_{Omh^2(z_i,z_j)_{obs}}}\right]^2
\end{align}
Then, we minimize this $\chi^2$ function to find the best-fitted
cosmological parameters.

\section{Results and Discussion}\label{sec:reanddi}

Let us discuss the results starting with $\Lambda$CDM model.
Numerical details are displayed in Table.~(\ref{table1}) and
comprise fits of $\Omega_{m,0}$ on different sub-samples using three
techniques: reduced chi-square Eq.(\ref{eqchi_Hmin}), chi-square
with Gaussian priors on $H_0$ and $Omh^2$ two-point diagnostics.
Full data-set without the high-$z$ BAO point (i.e. $n=35$ data
points), gives $\Omega_{m,0}=0.30^{+0.10}_{-0.07}$ and
$H_0=67.55^{+4.57}_{-4.33}$ km/s/Mpc, whereas homogeneous DA sample
($n=30$ data points) results with
$\Omega_{m,0}=0.32^{+0.10}_{-0.08}$ and $H_0=67.74^{+4.95}_{-4.37}$
km/s/Mpc, respectively. These two central values are very close to
\textit{Planck} central fits result \citep{Planck2014}, that is
$\Omega_{m,0}=0.314\pm0.020$ and $H_0=67.4\pm1.4\;\;km\, s^{-1}\,
Mpc^{-1}$. When the high-$z$ point included in the sub-sample, the
results changes to $\Omega_{m,0}=0.28^{+0.09}_{-0.07}$ and
$\Omega_{m,0}=0.26^{+0.05}_{-0.04}$ for the 6 H(z) from BAO and the
whole H(z) sample, respectively. And corresponding Hubble constant
are $H_0=66.04^{+7.89}_{-6.86}$ km/s/Mpc and
$H_0=69.14^{+3.75}_{-3.54}$ km/s/Mpc, respectively. The result is
still consistent with Planck result in $1\sigma$ but there is a
mismatch in central fits. Results obtained with priors on $H_0$
reveal that $\Omega_{m,0}$ fit is sensitive to the Hubble constant
assumed. Inclusion of the $H(z=2.34)$ point leads to results which
are biased with respect to \textit{Planck} result. One can also see
that $Omh^2(z_i,z_j)$ two-point diagnostics gives more stringent
results than $H(z)$ alone.

\begin{table*}[htp]
\caption{Best fitted parameters in wCDM cosmological model using
$H(z)$ data alone and $Omh^2(z_i,z_j)$ two-point diagnostics. Fits
done on different sub-samples are reported. First panel corresponds
to the reduced $\chi^2$ method. Second and third panels corresponds
to priors on $H_0$ taken after \textit{Planck} \citep{Planck2014}
and after \citet{Riess2016}. }
\begin{center}
{{\scriptsize
 \begin{tabular}{l c c c } \hline\hline
 $\chi^2_{H(z)}(\textbf{p})$  & $\Omega_{m,0}$  & $w$ &  $\chi^2_{d.o.f}$ \\ \cline{2-4}
 reduced chi-square & & & \\
$n=6$ BAO H(z)  & $0.26^{+0.15}_{-0.18}$ & $-0.37^{+0.11}_{-2.07}$ & $0.48/4$ \\
$n=30$ DA H(z)  & $0.29^{+0.13}_{-0.09}$ & $-1.21^{+0.69}_{-1.40}$ &  $14.36/28$ \\
$n=35$ H(z)     & $0.28^{+0.10}_{-0.05}$ & $-1.53^{+0.70}_{-1.06}$ &  $15.77/33$ \\
$n=36$ all H(z) & $0.26^{+0.06}_{-0.04}$ & $-0.95^{+0.40}_{-0.49}$ &  $17.76/34$ \\
 \hline \hline
  $\mathcal{L}_H(\textbf{p})$ & $\Omega_{m,0}$  & $w$ & $\chi^2_{d.o.f}$ \\ \cline{2-4}
  Gaussian Prior & $H_0=67.4\pm1.4 $ km/s/Mpc & & \\
$n=6$ BAO H(z)  & $0.28^{+0.04}_{-0.05}$ & $-1.04^{+0.28}_{-0.34}$ &  $1.19/4$ \\
$n=30$ DA H(z)  & $0.34^{+0.08}_{-0.13}$ & $-0.97^{+0.36}_{-0.55}$ & $15.14/28$ \\
$n=35$ H(z)     & $0.34^{+0.07}_{-0.11}$ & $-1.08^{+0.31}_{-0.49}$ & $17.43/33$\\
$n=36$ all H(z) & $0.27^{+0.04}_{-0.05}$ & $-0.90^{+0.20}_{-0.25}$ &
$18.74/34$\\ \cline{2-4}
Gaussian Prior  & $H_0=73.24\pm1.74 $ km/s/Mpc & & \\
$n=6$ BAO H(z)  & $0.24^{+0.03}_{-0.04}$ & $-1.18^{+0.24}_{-0.38}$ &  $1.20/4$\\
$n=30$ DA H(z)  & $0.32^{+0.05}_{-0.09}$ & $-1.40^{+0.45}_{-0.56}$ &  $15.16/28$\\
$n=35$ H(z)     & $0.31^{+0.05}_{-0.07}$ & $-1.45^{+0.41}_{-0.53}$ &  $16.70/33$\\
$n=36$ all H(z) & $0.25^{+0.03}_{-0.03}$ & $-1.15^{+0.22}_{-0.25}$ &  $19.25/34$\\
 \hline \hline
$\chi^2_{Omh^2}(H_0,\textbf{p})$ & $\Omega_{m,0}$  & $w$  &
$\chi^2_{d.o.f}$ \\ \cline{2-4}
Uniform Prior &  $H_0\in[66.0, 68.8]$ km/s/Mpc  & & \\
$n=6$ BAO H(z)  & $0.28^{+0.01}_{-0.03}$ & $<-0.82$  &  $1.18/12$ \\
$n=30$ DA H(z)  & $0.30^{+0.03}_{-0.10}$ & $-0.80^{+0.22}_{-0.32}$ &  $217.08/432$\\
$n=35$ H(z)     & $0.33^{+0.02}_{-0.03}$ & $-1.20^{+0.26}_{-0.40}$ &  $292.35/592$ \\
$n=36$ all H(z) & $0.25^{+0.02}_{-0.02}$ & $-0.78^{+0.11}_{-0.11}$ &
$312.82/627$ \\  \cline{2-4}
Uniform Prior &  $H_0\in[71.50, 74.98]$ km/s/Mpc  & & \\
$n=6$ BAO H(z)  & $0.24^{+0.01}_{-0.02}$ & $<-0.88$                &  $1.16/12$\\
$n=30$ DA H(z)  & $0.26^{+0.02}_{-0.07}$ & $-0.85^{+0.18}_{-0.24}$ &  $216.51/432$\\
$n=35$ H(z)     & $0.28^{+0.02}_{-0.03}$ & $-1.14^{+0.19}_{-0.28}$ &  $291.56/592$\\
$n=36$ all H(z) & $0.22^{+0.01}_{-0.02}$ & $-0.82^{+0.09}_{-0.09}$ &  $312.38/627$\\
 \hline \hline
\end{tabular}} \label{table2}}
\end{center}
\end{table*}

\begin{table*}[htp]
\caption{Best fitted parameters in CPL cosmological model using
$H(z)$ data alone and $Omh^2(z_i,z_j)$ two-point diagnostics. Fits
done on differential ages of cosmic chronometers (DA) and on the
full sample enriched with BAO data. Priors on $H_0$ were taken after
\textit{Planck} \citep{Planck2014} and after \citet{Riess2016}. }
\begin{center}
{{\scriptsize
 \begin{tabular}{l c c c c c} \hline\hline
 $\chi^2_{H(z)}(\textbf{p})$ & $\Omega_{m,0}$  & $w_0$  & $w_a$ & $\chi^2_{d.o.f}$ \\ \cline{2-5}
 reduced chi-square  & & & & \\
$n=30$ DA H(z) & $0.40^{+0.42}_{-0.19}$ & $-0.35^{+5.19}_{-2.20}$ & $-0.65^{+4.11}_{-35.23}$  & $14.28/27$\\
$n=36$ all H(z) & $0.27^{+0.17}_{-0.14}$ & $-0.91^{+1.14}_{-0.82}$ & $0.73^{+1.75}_{-4.08}$  & $17.29/33$\\
 \hline \hline
 $\mathcal{L}_H(\textbf{p})$  & $\Omega_{m,0}$  & $w_0$  & $w_a$ & $\chi^2_{d.o.f}$ \\ \cline{2-5}
    Gaussian Prior & $H_0=67.4\pm1.4 $ km/s/Mpc & & &\\
$n=30$ DA H(z) & $0.41^{+0.06}_{-0.10}$ & $-0.76^{+1.17}_{-0.82}$ & $0.05^{+1.93}_{-15.90}$ & $15.12/27$\\
$n=36$ all H(z) & $0.30^{+0.05}_{-0.11}$ & $-0.84^{+0.35}_{-0.38}$ & $0.73^{+0.66}_{-3.37}$ & $18.50/33$\\
\cline{2-5}
Gaussian Prior  & $H_0=73.24\pm1.74 $ km/s/Mpc & & & \\
$n=30$ DA H(z) & $0.35^{+0.05}_{-0.10}$ & $-1.27^{+0.91}_{-0.96}$ & $1.30^{+1.99}_{-13.92}$ & $15.00/27$ \\
$n=36$ all H(z) & $0.25^{+0.04}_{-0.09}$ & $-1.17^{+0.30}_{-0.40}$ & $1.17^{+0.71}_{-2.41}$ & $18.54/33$\\
 \hline \hline

   $\chi^2_{Omh^2}(H_0,\textbf{p})$  & $\Omega_{m,0}$  & $w_0$ & $w_a$ & $\chi^2_{d.o.f}$ \\ \cline{2-5}
Uniform Prior & $H_0\in[66.0, 68.8]$ km/s/Mpc  & & & \\
$n=30$ DA H(z) & $0.29^{+0.06}_{-0.16}$ & $-0.80^{+0.24}_{-0.45}$ & $0.65^{+0.76}_{-0.97}$ & $217.37/431$ \\
$n=36$ all H(z) & $0.20^{+0.06}_{-0.11}$ & $-0.82^{+0.18}_{-0.17}$ & $0.71^{+0.26}_{-0.38}$ & $309.95/626$ \\
\cline{2-5}
Uniform Prior &  $H_0\in[71.50, 74.98]$ km/s/Mpc  & & \\
$n=30$ DA H(z) & $0.23^{+0.05}_{-0.19}$ & $-0.87^{+0.25}_{-0.33}$ & $0.74^{+0.57}_{-0.78}$ & $216.96/431$ \\
$n=36$ all H(z) & $0.17^{+0.04}_{-0.09}$ & $-0.88^{+0.15}_{-0.15}$ & $0.79^{+0.24}_{-0.34}$ & $308.82/626$ \\
 \hline \hline
\end{tabular}} \label{table3}}
\end{center}
\end{table*}

Results concerning wCDM model are reported in Table~(\ref{table2})
and shown on Fig.~(\ref{fig6}) -- Fig.~(\ref{fig7}). When we take
the prior $H_0=67.4\pm2.4$ km/s/Mpc from \cite{Planck2014}, the dark
energy equation of state constraint is almost totally consistent
with $\Lambda$CDM where $w=-1$. However, the prior
$H_0=73.24\pm1.74$ km/s/Mpc from \citet{Riess2011}, favors Phantom
behavior ($w<-1$). The conclusion is that fits are very sensitive to
the value of $H_0$, which is consistent with findings of
\citet{Farooq2013}. When we use $H(z)$ measurements from BAO and DA
techniques separately, the results are different: $H(z)$ data from
DA favor quintessence($w>-1$) while $H(z)$ data from BAO favor
phantom ($w<-1$) fields. Moreover the $H(z=2.34)$ point has a big
leverage on the final results. This is consistent with conclusions
of our previous works \citep{Zheng16,Ding2015}. The reason lies in
different systematic effects between BAO and DA. Better restrictive
power of $Omh^2(z_i,z_j)$ as compared with $H(z)$ technique can be
understood in terms of the sample size. Namely, a sample of $n$
$H(z)$ measurements provides us with $\frac{n(n-1)}{2}$
$Omh^2(z_i,z_j)$ data-points. This advantage does not show up for
small samples like $n=6$ BAO $H(z)$ data-points. However, the
$Omh^2(z_i,z_j)$ diagnostic have a certain drawback: because of
$H_0$ is strongly degenerated with other cosmological parameters, it
should be better to give a prior $H_0$ value.

Finally the results concerning CPL parametrization are shown in
Table~(\ref{table3}) and on Fig.~(\ref{fig8}) for 30 DA $H(z)$ and
the whole 36 $H(z)$ sample. The $\Lambda$CDM model in which $w_0=-1$
and $w_a=0$ is identified in Fig.~(\ref{fig8})by a black star. The
$Omh^2(z_i,z_j)$ diagnostics provides much more stringent results
for the dark energy equation of state. For the 30 DA $H(z)$, i.e.
the homogeneous sample of cosmic chronometers the black star
indicating $\Lambda$CDM model stays at the edge of 1$\sigma$
confidence region, while it is outside this region for the full,
mixed sample of $n=36$ data points. This illustrates the
aforementioned systematic effects associated with BAO measurements.

\section{Conclusion}\label{sec:conclusion}

With increasing number of cosmic chronometers
\citep{Moresco2015,Moresco2016a} covering bigger redshift range, we
are starting to directly probe the expansion of the Universe through
measurements of its expansion rates $H(z)$ at different epochs. More
importantly this sort of measurements is not entangled with cosmic
distance ladder considerations or any other calibrations
pre-assuming cosmological model. However, there have been some
misunderstanding in this respect since additional measurements of
$H(z)$ from BAO peaks location were used in the literature as well.
In order to discuss this issue and show the performance of $H(z)$
data in the context of cosmological model testing, we used recently
most complete, mixed data coming from differential ages of passively
evolving galaxies together with BAO data-points. Besides such full,
inhomogeneous data-set we considered homogeneous sub-samples as
well. One of the conclusions was that BAO and DA data should not be
mixed together for the purpose of testing cosmological models. This
can be understood because BAO technique pre-assumes cosmological
model in order to disentangle BAO peak position from the
redshift-space distortions due to peculiar velocities of galaxies.
Indications of a bias introduced by BAO data has also been noticed
in \citet{Zheng16}.

In this paper we used both pure expansion rates $H(z)$ and two point
diagnostics $Omh^2(z_i, z_j)$. The latter has originally been
invoked as a litmus test for the $\Lambda$CDM. There were ideas for
using it in broader context \citep{Sahni2008} illustrated with
simulated future data. Here, we applied the two-point diagnostics on
the real data and demonstrated that they are able to give much
stringent constraints on cosmological parameters. This is because of
enhanced size of the data-set: from $n$ original $H(z)$ data-points
one can get $n(n-1)/2$ two-point diagnostics. The price one pays is
that they are strongly correlated. Let us stress that the chi-square
function we used was not meant to follow the chi-square
distribution, but it only served a purpose to define the likelihood
function to be maximized with MCMC simulations. Even though the
constraining power of $Omh^2(z_i, z_j)$ two-point diagnostics is
considerable, it suffers from being sensitive to the $H_0$ prior.
Hence the performance of this method crucially depends on our
knowledge about the correct value of the Hubble constant. When this
work has been completed, \citet{Melia2017} published an important
paper in which they introduced a new type of two-point diagnostics,
completely independent on the Hubble constant $H_0$. They also gave
much more rigorous treatment of statistical properties of this
diagnostics. It would be interesting to use their approach in a
similar way we did in this paper.

\section*{Acknowledgements}

This work was supported by National Key Research and Development 
Program of China No. 2017YFA0402600, the National Basic Science 
Program (Project 973) of China under (Grant No. 2014CB845800), 
the National Natural Science Foundation of China under Grants 
Nos. 11503001, 11690023, 11373014, and 11633001, the Strategic 
Priority Research Program of the Chinese Academy of Sciences, 
Grant No. XDB23000000, the Interdiscipline Research Funds of 
Beijing Normal University, and the Opening Project of Key 
Laboratory of Computational Astrophysics, National Astronomical 
Observatories, Chinese Academy of Sciences. This research was 
also partly supported by the Poland-China Scientific \& 
Technological Cooperation Committee Project No. 35-4. M.B. was 
supported by Foreign Talent Introducing Project and Special 
Fund Support of Foreign Knowledge Introducing Project in China.


\label{lastpage}

\maketitle


\begin{thebibliography}{99}
\bibitem[Riess et~al.(1998)]{Riess1998} A. G. Riess et al., 1998, Astron. J. 116, 1009 (1998)
\bibitem[Perlmutter et~al.(1999)]{Perlmutter1999} S. Perlmutter et al., Astrophys. J. 517, 565 (1999)
\bibitem[Clifton et~al.(2012)]{Clifton2012} T. Clifton, P. G. Ferreira, A. Padilla, and C. Skordis, Physics Reports 513, 1 (2012)
\bibitem[Frieman et~al.(2008)]{Frieman2008} J. Frieman, M. Turner, and D. Huterer, Ann. Rev. Astron. Astrophys. 46, 385-432 (2008)
\bibitem[Buchert et~al.(2016)]{Buchert16} T. Buchert, A. A. Coley, H. Kleinert, B. F. Roukema, and D. L. Wiltshire, Int. J. Mod. Phys. D 25, 1630007 (2016)
\bibitem[Chevalier \& Polarski (2001)]{Chevalier2001} M. Chevalier, and D. Polarski, Int. J. Mod. Phys. D 10, 213-224 (2001)
\bibitem[Linder (2003)]{Linder2003} E. V. Linder, Phys. Rev. Lett. 90, 091301 (2003)
\bibitem[\protect\citeauthoryear{Cao et~al.}{2012}]{Cao12} S. Cao, et al. JCAP, 03, 016 (2012)
\bibitem[\protect\citeauthoryear{Cao et~al.}{2014}]{Cao14} S. Cao, et al. PhRvD, 90, 083006 (2014)
\bibitem[\protect\citeauthoryear{Cao et~al.}{2015}]{Cao15} S. Cao, et al. ApJ, 806, 185 (2015)
\bibitem[\protect\citeauthoryear{Cao et~al.}{2017a}]{Cao17a} S. Cao, et al. JCAP, 02, 012 (2017)
\bibitem[\protect\citeauthoryear{Cao et~al.}{2017b}]{Cao17b} S. Cao, et al. A\&A, 606, A15 (2017)
\bibitem[Moresco et~al.(2015)]{Moresco2015} M. Moresco, Mon. Not. R. Astro. Soc. 450, L16 (2015)
\bibitem[Moresco et~al.(2016a)]{Moresco2016a} M. Moresco et al., JCAP, 05, 014 (2016)
\bibitem[Ding et~al.(2015)]{Ding2015} X. Ding, M. Biesiada, S. Cao, Z. X. Li, and Z. H. Zhu, Astrophys. J. Lett. 803, L22 (2015)
\bibitem[Moresco et~al.(2016b)]{Moresco2016b} M. Moresco et al., JCAP, 12, 029 (2016)
\bibitem[Sahni, Shafieloo \& Starobinsky (2008)]{Sahni2008} V. Sahni, A. Shafieloo, and A. A. Starobinsky, Phys. Rev. D 78, 103502 (2008)
\bibitem[Shafieloo, Sahni \& Starobinsky (2012)]{Sahni2012} A. Shafieloo, V. Sahni, and A. A. Starobinsky, Phys. Rev. D 86, 103527 (2012)
\bibitem[Sahni, Shafieloo \& Starobinsky (2014)]{Sahni2014} V. Sahni, A. Shafieloo, and A. A. Starobinsky, Astrophys. J. Lett. 793, L40 (2014)
\bibitem[Ade et~al.(2014)]{Planck2014} P. A. R. Ade et al., [Planck Collaboration] Astron. Astrophys. 571, A16 (2014)
\bibitem[Zheng et~al.(2016)]{Zheng16} X. G. Zheng, X. H. Ding, M. Biesiada, S. Cao, and Z. H. Zhu, Astrophys. J. 825, 17 (2016)
\bibitem[\protect\citeauthoryear{Cao et~al.}{2011a}]{Cao11a} S. Cao, et al.  A\&A, 529, A61 (2011)
\bibitem[\protect\citeauthoryear{Cao et~al.}{2011b}]{Cao11b} S. Cao, et al. MNRAS, 416, 1099 (2011)
\bibitem[\protect\citeauthoryear{Cao et~al.}{2013}]{Cao13} S. Cao, et al. IJMPD, 22, 1350082 (2013)
\bibitem[\protect\citeauthoryear{Chen et~al.}{2015}]{Chen15} Y. Chen, et al. JCAP, 02, 010 (2015)
\bibitem[Jimenez et~al.(2003)]{Jimenez2003} R. Jimenez, L. Verde, T. Treu, and D. Stern, Astrophys. J. 593, 622 (2003)
\bibitem[Zhang et~al.(2014)]{Zhang2014} C. Zhang et al., Res. Astron. Astrophys. 14, 1221 (2014)
\bibitem[Moresco et~al.(2012)]{Moresco2012} M. Moresco et al., JCAP, 08 (2012) 006
\bibitem[Simon et~al.(2005)]{Simon2005} J. Simon, L. Verde, and R. Jimenez, Phys. Rev. D 71, 123001 (2005)
\bibitem[Stern et~al.(2010)]{Stern2010} D. Stern, et al. JCAP, 02, 008 (2010)
\bibitem[Anderson et~al.(2013)]{Anderson2013} L. Anderson et al., Mon. Not. R. Astro. Soc. 439, 83 (2014)
\bibitem[Blake et~al.(2012)]{Blake2012} C. Blake et al., Mon. Not. R. Astro. Soc. 425, 405 (2012)
\bibitem[Chuang et~al.(2013)]{Chuang2013} C. H. Chuang, and Y. Wang, Mon. Not. R. Astro. Soc. 435, 255-262 (2013)
\bibitem[Delubac et~al.(2015)]{Delubac2015} T. Delubac et al., Astron. Astrophys. 574, A59 (2015)
\bibitem[Farooq(2013)]{Farooq2013} M. O. Farooq, arXiv:1309.3710
\bibitem[Riess et~al.(2011)]{Riess2011} A. G. Riess et al., Astrophys. J. 730, 119 (2011)
\bibitem[Riess et~al.(2016)]{Riess2016} A. G. Riess et al., Astrophys. J. 826, 56 (2016)
\bibitem[Leaf \& Melia (2017)]{Melia2017} K. Leaf, and F. Melia, Mon. Not. R. Astro. Soc. 470, 2320 (2017)
\end{thebibliography}
\end{document}